\documentclass[aps,prl,twocolumn]{revtex4}
\usepackage{color}
\usepackage{graphicx}
\usepackage{dcolumn}
\begin{document}

\title{Improved selectivity of optical transmission through cascaded waveguide-metal-grating filters}

\author{J\'er\^ome Le Perchec}\email{jerome.le-perchec@cea.fr}

\affiliation{CEA, LETI, Minatec Campus, Optics and Photonics Department, 17 avenue des Martyrs, 38054 Grenoble, France}

\date{\today}
\begin{abstract} 
The resonant transmission of two near-field coupled, cascaded band-pass filters, based on metallic stripe or patch gratings, is analysed. The response, in terms both of maximum efficiency and light rejection out of the resonance, overpasses the simple convolution of the responses of two isolated filters, while keeping good angular tolerance. Illustrations are given in the infrared range where sizing and technological integration of such compact structures are particularly relevant for detection applications.
\end{abstract}


\maketitle
Guided-mode resonant (GMR) filters are based on the electromagnetic resonance of a waveguide mode inside a high index dielectric slab coupled to a sub-wavelength grating. Purely dielectric structures allow getting extremely narrow transmission or reflection peaks \cite{Tibuleac2001,Jacob2002,Kawanishi2020}, for highly wavelength-selective applications going from infrared (IR) to visible ranges. Their sharp spectral responses are generally very sensitive to the incidence angle and to the finite size of the grating  \cite{Bendickson}. When the grating consists of thin \textit{metallic} two-dimensional (2D) patches (or infinite stripes), broader and high transmission peaks may be easily obtained, less sensitive to the incidence angle and/or light polarization \cite{Leperchec2011,Kaplan2011}. In some extent, such a kind of structure joins the well-known family of the frequency-selective surfaces (metal meshes) with, here, a dielectric loading \cite{Luebbers1978}. The compactness of GMR filters and the resonance tunability with the lateral dimensions of the grating are particularly advantageous compared to thicker multi-layered structures, whose efficiency and angular-tolerance are also interesting \cite{Im2017,Frey2018}.

\begin{figure}
\begin{center}
\includegraphics[angle=0,scale=.55]{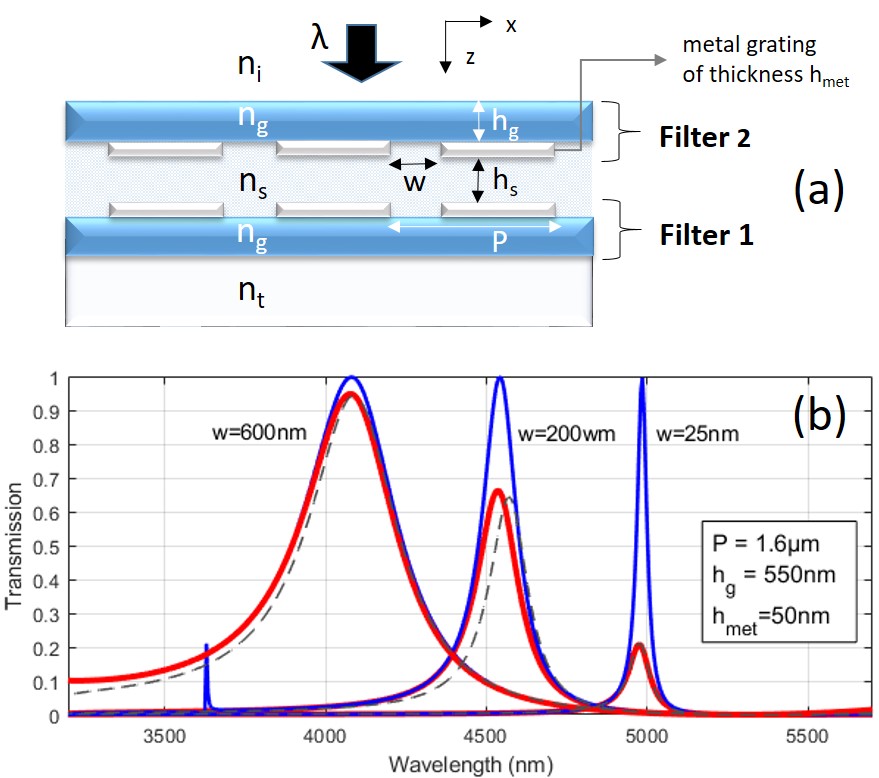}
\end{center}
\caption{(a) Sketch of two cascaded GMR filters with a spacer layer of low index $n_s$ between them. (b) Reference transmission of the \textit{isolated} band-pass filter number 1 (only air above it), by taking $n_g=n_{Ge}=4$, $n_{met}=n_{Alu}$, and $n_{t}$=1 (1D stripe grating, TM-polarization). RCWA simulations are given for three aperture widths $w$ and compare the real metal case (red curves, complex metal permittivity $\varepsilon_{Alu}$) to a purely reflecting but non-absorbing case by simply omitting the imaginary part of $\varepsilon_{Alu}$ (blue lines). For instance, at $\lambda=4\mu$m, $\varepsilon_{Alu}=-1543+512i$. Dotted curves result from the simplified analytical model (see Appendix) with the complex permittivity. The fine peak at $\lambda=3.6\mu$m is a resonance of metallic antennas that does not exist with real metals}
\label{figure1}
\end{figure}

In case of high-performance optical detectors with a somewhat large field of view, like quantum detectors used for IR imaging, the working spectral range  may be rather large (for instance $[3-5.5]\mu$m wavelength for mid-wave IR applications). Thus, band-pass GMR filters should present a high transmission peak (at least $>70\%$) around a wavelength of interest, with very low sideband transmittance ($<0.1\%$ preferred) over the rest of the spectrum. A single metallic-grating/waveguide filter may not offer sufficient light rejection out of resonance. Thus, in the present work, we study the selective transmission properties offered by the near-field stacking of two similar GMR filters (see Fig.\ref{figure1}(a)) like the ones previously studied \cite{Leperchec2011}. A quite similar architecture was already implemented in double-layer metallic subwavelength slit arrays \cite{Chan2006}, but not based on filters with high-index slabs (not the same mechanisms). We aim at getting \textit{significantly improved light rejection} out of the pass-band without altering the maximum peak. Very sharp resonances (like the ones we may find with coupled dielectric gratings \cite{Song2009}) are not necessarily sought here as some applications like hyperspectral imgaging may need to use some part of the spectral window around a wavelength of interest. 
We mainly investigate the case of cascaded filters with realistic materials and geometrical parameters, designed for the mid-infrared, using Rigorous Coupled Wave Analysis (RCWA) in 1D. The 2D configuration is also discussed at the end of this paper. Before tackling the cascaded structure, we describe the essential behavior of the single filter with the support of practical analytical formulae.

In a 1D configuration, the resonant mode of interest is found under TM-polarization (incident magnetic field colinear to the metallic stripes). Other transmission peaks could be observed in TE-polarization, but generally show less interesting spectral profiles. Simple design rules of such GMR filters have already been given \cite{Leperchec2011}. We remind that the waveguide is advantageously of high optical index $n_g$ in order to get a better resonance quality and an acceptable robustness with respect to the incidence angle $\theta$ ($\pm 10^{o}$ typically, which might be increased by resorting to a bi-atomic structuration\cite{Sakat2013}). Its thickness $h_g$ is preferentially $\sim\lambda_0/2$, where $\lambda_0$ is the wanted resonance wavelength. The periodicity $P$ of the thin grating is such that $\lambda_0/n_g <P<2 \lambda_0/n_g$ in order to have a propagating first-order mode inside the slab. Figure\ref{figure1}(b) gives a reference example taking a Germanium/Aluminum structure surrounded by air. It clearly shows that, at given period, the slit width $w$ should be narrow if we seek spectral selectivity, and allows tuning the transmission amplitude and bandwidth. However, it may be to the detriment of the efficiency because of optical absorption in the metal.

A way to analytically calculate the transmission of a single filter has been given in a previous work \cite{Leperchec2011}, based on a simplified modal method, for highly reflecting metallic-stripe grating. We give in the Appendix an even more general expression with arbitrary optical indices $n_i,n_g,n_d,n_t$ of the incidence, intra-slit, waveguide, and transmission regions, respectively. This formula gives reliable quantitative results compared to RCWA simulations, as shown in Fig.\ref{figure1}(b). Moreover, by tending to the perfect metal, new key relations were found and presented hereafter, in particular the explicit resonance condition of such filters. While $kh_{met}<<1$ ($k=2\pi/\lambda$), at $\theta=0$, the TM-transmission reads:
\begin{widetext}
\begin{equation}
 Tr \approx \frac{n_{i}}{n_{t}} \left|\frac{ 2}{\cos(kn_g h_g) \left[\left(1+\frac{n_{i}}{n_t}\right)+\frac{2}{n_t}\sum_{m >0}A_m \right]- j\sin(kn_g h_g) \left[\left(\frac{n_g}{n_t}+\frac{n_{i}}{n_g}\right)+\frac{2}{n_g}\sum_{m >0}A_m\right] }\right|^2
\label{eq:transm}\end{equation}
\end{widetext}
where $j$ is such that $j^2=-1$ and 
\begin{equation}
A_m(\lambda)=\sec^2(m\pi\frac{w}{P}) \left[\frac{n_i}{\beta_m^{i}}+\frac{n_g}{\beta_m^g} \left( \frac{1-j Y_m\tan X_m}{ Y_m-j\tan X_m}\right) \right]
\label{eq:Am}\end{equation}
with $X_m=k n_g h_g \beta_m^g$ and $\beta_m^g=\sqrt{1-(m\lambda/(n_g P))^2}$, 
m being a relative integer (for $\beta_m^t$, replace $n_g$ by $n_t$). The terms $k\beta_m$ are wave vectors along the z-direction. 
$Y_m=(n_g \beta_m^{t})/(n_{t} \beta_m^g)$ are Floquet modal impedance ratios. We remark that taking $w=P$, all $A_m=0$ and we retrieve the transmission of a homogeneous slab of index $n_g$. 
When $\lambda<n_g P$ ($\lambda<6.4\mu$m in the above example), the first-order mode inside the waveguide is propagating, i.e. $\beta_1^g>0$ and $A_1$ becomes purely imaginary. We thus find the first transmission resonance when the imaginary part of the denominator in (\ref{eq:transm}) is zero, say:
\begin{equation}
\left(\frac{n_g}{n_t}+\frac{n_{i}}{n_g}\right)\tan (kn_g h_g) \approx \frac{2}{n_t} \Im(A_1)
\label{eq:rescond}
\end{equation}
by willingly neglecting high-order diffraction terms ($m\geq 2$). The resonance condition is actually more precise by taking at least $\Im(A_1+A_2)$, which gives $\lambda_0=4.04\mu$m when $w=600$nm in our example. Finally, given (\ref{eq:rescond}), the transmission peak is:
\begin{equation}
Tr_{\lambda=\lambda_0} \approx \frac{n_{i}}{n_{t}} \left|\frac{2\cos (k_0n_g h_g)}{1+\frac{n_{i}}{n_{t}}\cos^2(k_0kn_g h_g) +\frac{n_i n_t}{n_g^2}\sin^2(k_0n_g h_g)}\right|^2
\end{equation}
Contrary to what could be thought, the aperture ratio $w/P$ has no impact on this amplitude (which is maximized when $\lambda_0/(2n_g) \sim h_g$). This is also true if the metal were simply not absorbing (purely negative permittivity), as shown in Fig.\ref{figure1}(b) (blue curve). $w/P$ has actually an impact on (\ref{eq:Am}) and therofore on the dispersion relation (\ref{eq:rescond}): when it diminishes, it makes  $\lambda_0$ shift to lower frequencies, whereas the resonance quality factor increases. But, as the transmission peak becomes sharper, it is more attenuated in a real metal case because of optical absorption.

Since we cannot overcome the natural trade-off between a high transmission peak (large ratio $w/P$) and a strong light rejection out of the resonance, we investigate now a cascaded structure where two identical filters are closely superposed (see Fig.\ref{figure1}(a)), with fairly large opening widths $w$. The spacer layer separating both filters is characterized by a low optical index in order prevent or minimize diffraction perturbation ($n_s\leq 1.5$, by using e.g. $SiO_2$, $MgF_2$, $YF_3$) and by a sub-wavelength thickness $h_s$ (further explanation given afterwards).

\begin{figure}
\begin{center}
\includegraphics[angle=0,scale=.50]{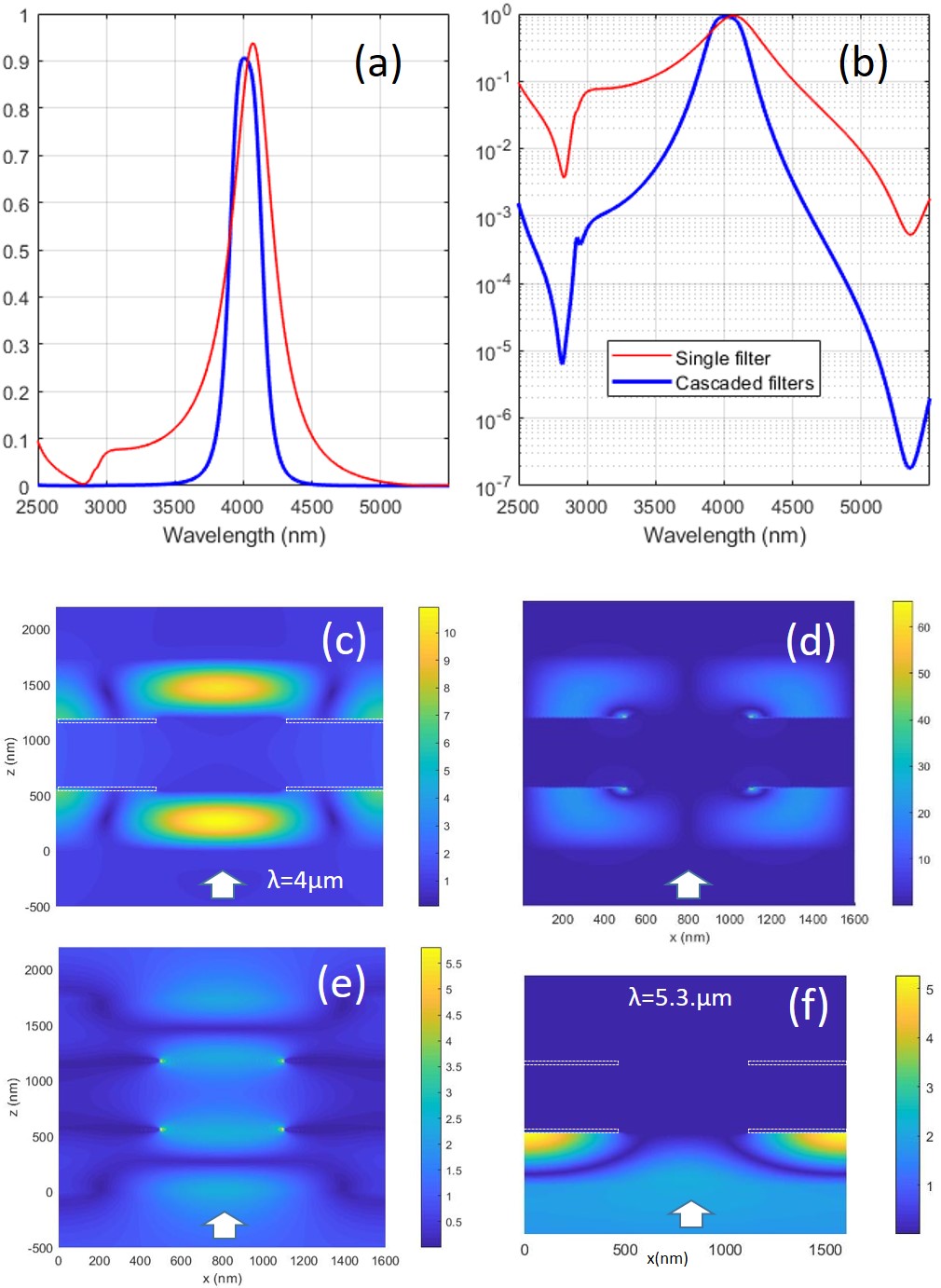}
\end{center}
\caption{(top) (a) Transmission spectra of two cascaded same Ge/Alu filters as sketched in Fig.\ref{figure1}(a), compared to that the single filter (surrounded by air), with $P=1.6\mu$m, $w=600$nm, $h_g=550$nm, $h_{met}=20$nm, $n_s=1$ and $h_s=600$nm. (b) Same transmission in logarithmic scale. (bottom) Spatial maps of the normalized magnetic and electric fields at the resonance wavelength $\lambda_0$: (c) $|H_y|$, (d) $|E_x|$, and (e) $|D_z|=|\varepsilon (x,z)E_z|$. Here, incident light comes from $z<0$, and the slit aperture is centered at $x=800$nm. (f) Cartography of $H_y$ when a transmission null occurs. }
\label{figure2}
\end{figure}

Figure \ref{figure2}(a) shows the transmission profile on the wide interval $[2.5;5.5]\mu$m, resulting from the cascaded structure based on our reference filters with $w=600$nm and $h_{met}$=20nm only, separated by an air layer of thickness $h_s=600$nm. The spectral response is now much cleaner and selective: light rejection is immediately reinforced, without deteriorating the maximum transmission efficiency. The resonance bandwidth is a bit tightened and a slight shift of $\lambda_0$ is observed, from $4.07$ to $4.01\mu$m. The responses in logarithmic scale (Fig.\ref{figure2}(b)) show that, surprisingly, rejection is better than the square of the rejection of isolated filters (see at $\lambda=3$ or $4.5\mu$m, for instance), whereas the resonance peak is barely lowered ($90.6\%$ versus $93.8\%$ with the single filter). Thus, the far-field does not derives from a simple convolution of isolated filters, because of some near-field interaction (we come back to this point later). 
Fig.\ref{figure2}(c,d,e) gives maps of the electromagnetic fields ($H_y$, $E_x$, $D_z$) at the resonance, which are consistent with the excitation of the first-order hybrid waveguide mode, and shows highly symmetrical patterns, with light enhancements inside the waveguides and/or within slits (the latter present an effective dipole momentum, supported by $E_x$, with hot spots at the corners, which is intrinsic to the field scattered by such metallic apertures in TM-polarization \cite{Leperchec2015}). 

It is worth noticing that the resonance peak appears within a spectral window bounded by two transmission nulls, independent of $w$ and occurring when :
\begin{equation}
\tan(kn_g\beta_1^g h_g) +j\frac{Z+Y_1(\beta_1^g/n_g)}{\beta_1^g/n_g+ZY_1} = 0
\label{eq:zerotr}
\end{equation} 

where $Y_m$ is already given above, and $Z=1/\sqrt{\varepsilon_{metal}}$ (see Appendix). For instance, (\ref{eq:zerotr}) gives the wavelengths $\lambda=5382$ and $2816$nm: the filter exposed to incident light simply behaves as a good reflector with almost no absorption (no transmission and flat reflectivity $\sim 95\%$), and then blocks any energy transfer to the following filtering stage, as shown in Fig.\ref{figure2}(f).

\begin{figure}[h]
\begin{center}
\includegraphics[angle=0,scale=.62]{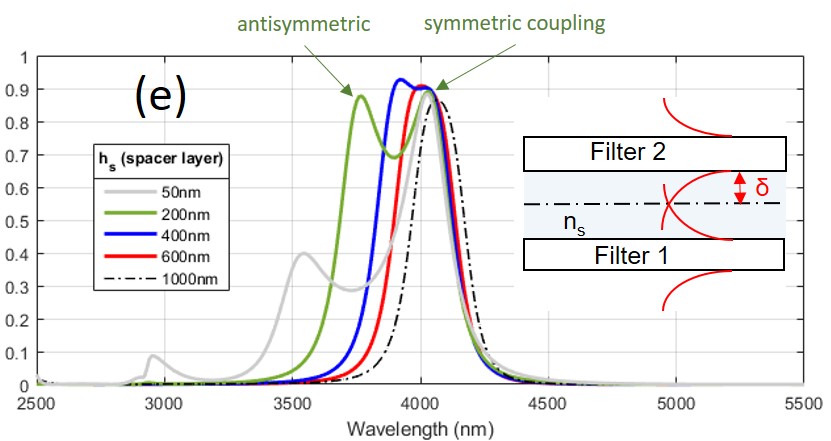}
\end{center}
\caption{Influence of the separation distance $h_s$. $\delta$ is the characteristic damping length of the evanescent 1rst order diffraction mode outside the waveguides.}\label{figure3}
\end{figure}

The influence of the spacer layer thickness ($h_s$) on the far-field response is illustrated in Fig.\ref{figure3}. When filters come closer, an additional part of energy may be transferred through the surrounding evanescent field, that is why the transmission peak is greater than the simple square $0.938^2$ (see the isolated filter).
Indeed, the resonating first order Rayleigh wave typically decreases as $exp(-kn_s|\beta^s_1 z|)=e^{-|z|/\delta}$
in the spacer layer, just outside the high-index waveguide, with $\delta =1/(n_sk_0\sqrt{(\lambda_0/(n_sP))^2-1})\approx 280$nm in our example. The best trade-off is when $h_s\sim 2\delta$.
Out of the resonance, the coupling plays an unfavorable role for the passage of light, hence a rejection better than expected. Below $300$nm separation, the near-field interaction becomes strong and creates a resonance splitting, with a symmetric and anti-symmetric coupling of filters. Here, antisymmetric coupling means that $\Re(E_x(x,z))$ changes of sign on both sides of the median plane separating both filters.

\begin{figure}[ht]
\begin{center}
\includegraphics[angle=0,scale=.48]{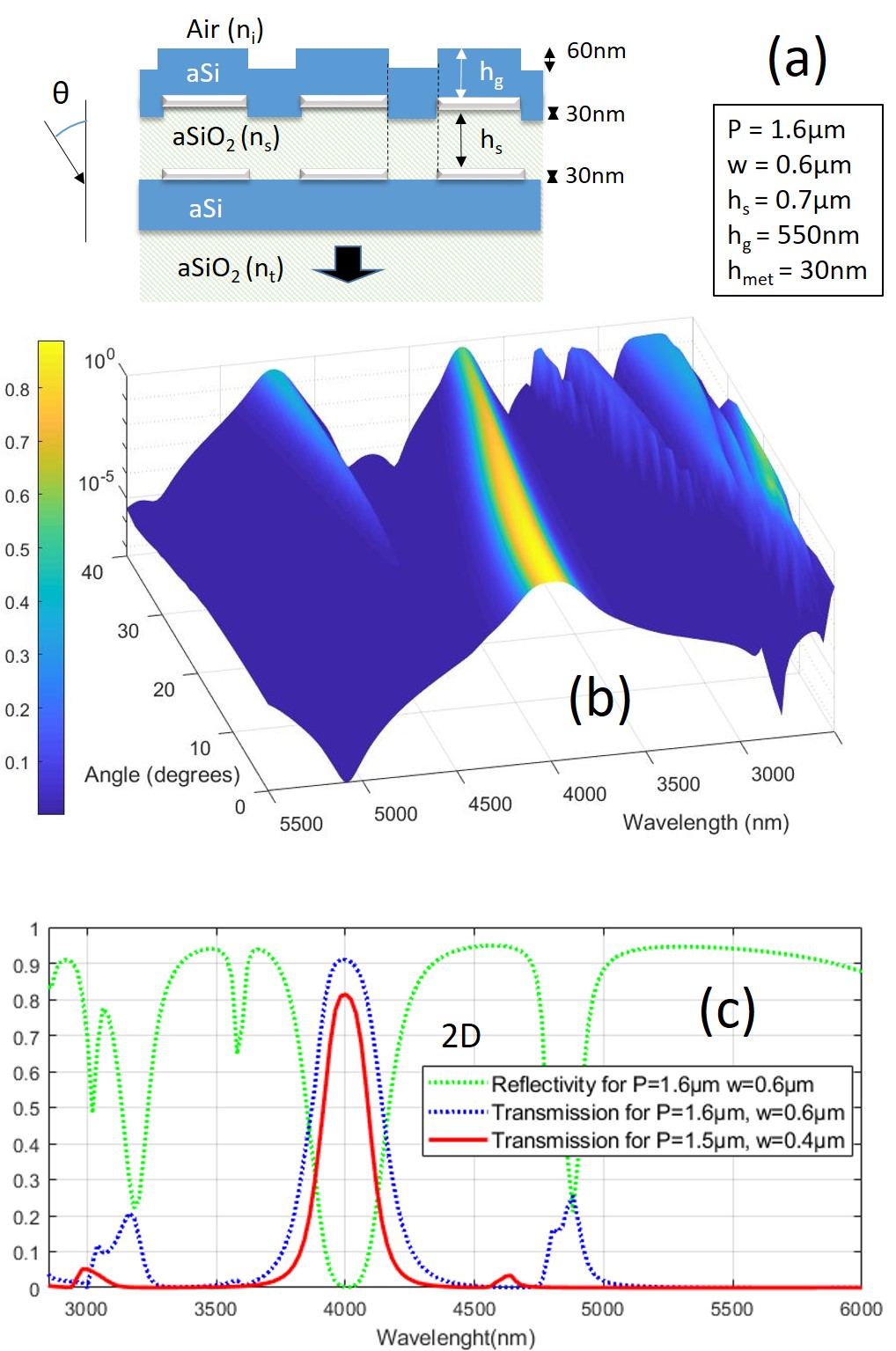}
\end{center}
\caption{(a) Possible perturbed filter architecture resulting from a CMOS technological process with a simple conformal deposition of successive layers once the first grating is made.  (b) Corresponding diagram of the transmission in function of $(\lambda,\theta)$. All materials permittivities taken from \cite{Palik} ($n_g\approx 3.84$ and $n_s=n_t\approx 1.4$). (c) 2D cascaded filters (square patch gratings), taking $n_g=4$, $n_{met}=n_{Alu}$, $n_i=n_t=n_s=1$, $h_g=550$nm, $h_{met}=40$nm, and $h_s=600$nm (to be compared to spectra of Fig.\ref{figure2}). With $n_s\sim 1.45$, the resonance peak would be slightly shifted towards $4.1\mu$m.} \label{figure4}
\end{figure}

The orientation of each grating with respect to the other (gratings above or under the waveguide, in an aligned stacking or a staggered arrangement) makes \textit{no} significant difference on the optical response with an ideal planar structure. However, realistic technological means may lead to a slightly perturbed geometry, for instance by using an auto-aligned technological process. Fig.\ref{figure4}(a) gives such an example with a CMOS-compatible technology, taking amorphous silicon-based materials. The diagram of the transmission as a function of $(\lambda,\theta)$ is given in log scale in Fig.\ref{figure4}(b). The maximum efficiency ($Tr_{max}= 88.8\%$) and rejection out of the resonance are still very good. If the grating of the upper filter 2 was put on the air side, the rejection would be a bit deteriorated at shorter wavelengths, and the transmission curve would lose in symmetry (not shown). Besides, the angular robustness (transmission level and inertia of the resonance wavelength) is typically $\theta=0\pm 15^{o}$, which is equivalent to that measured on single filters \cite{Leperchec2011}. At higher angles, an other waveguide mode (existing even in the perfect metal case) is excited around $\lambda=5\mu$m and is quite absorbing (contrary to the resonance at $\lambda=4\mu$m). For short wavelengths, new diffraction orders become propagating inside the high-index slab.

Geometrical differences between both filters might be willingly introduced to deform/tailor the transmission profile. The cascaded filter can stand alone in the air, or be bonded just above a detector. In the latter case, a low-index layer, whose thickness is adapted in the sense of constructive interference (quarter-wave plate or an odd multiple), must be inserted between the filter 1 and the detector underneath. If the detector is a matrix of pixels (e.g. with pixels of $20\mu$m in diameter), it may be useful to conveniently etch (trench) the cascaded stack along the pixels perimeter to avoid possible optical cross-talk through the waveguide. Since we are limited by the pixel area, a dozen of periods is necessary to get the wanted effect as we deal with a resonance that is not too sharp (out of the scope of this paper).

1D structures make it possible to discriminate the light polarization of the filtered signals. The conception of a polarization-independent 2D-structure (square patch grating) is accompanied with additional difficulties. This is pointed out through Fig.\ref{figure4}(c): whereas the high resonance peak of cascaded 2D-filters is still present at $\lambda=4\mu$m, secondary transmission peaks appear on both sides, which limit the spectral window where strong light rejection is wanted. These peaks actually correspond to waveguide modes we would get in $TE$-polarization with a 1D filter. These modes are also quite absorbing ones, contrary to the TM-resonance we work with (see reflectivity). A way to limit their excitation strength is to reduce $w$ (with some reduction of the period to retrieve the same $\lambda_0$), as the fundamental TE-slit mode is strongly evanescent when $\lambda \gg 2w$. Also, 2D metallic patches with rounded corners should be avoided in order to minimize the appearance of such perturbations \cite{Leperchec2011}. By reducing $w$, the main TM-resonance obviously loses in amplitude. Nevertheless, rejection levels remain strong: in the case $w=0.4\mu$m, $P=1.5\mu$m, $Tr\approx 10^{-3}$ when $\lambda\approx 3.5$ or $4.5\mu$m, and $Tr< 10^{-5}$ at $\lambda=5.5\mu$m. Another way to get an equivalent 2D-filter without TE-modes could be to resort to a polarization converter upstream to cascaded 1D-filters. Optical stages able to efficiently convert an entire signal into a given linear polarization state have already been demonstrated \cite{Kim2012}.

In conclusion, thanks to rigorous simulations and with the support of a modal approach, we have shown how two near-field coupled GMR filters allows obtaining a bandpass device with a transmission selectivity better than that expected by a simple convolution of isolated filters. Ideal and realistic examples have been especially given in the mid-infrared where such filtering properties (transmission peak $\sim 90\%$, strong rejection out of the resonance over a large spectral range, good angular tolerance) are demanded for hyper-spectral imaging and gas detection applications. Those structures, whose overall thickness remains smaller than the half-wavelength, are compact and rather easy to do on a technological level. 

\section{Appendix: Analytical calculation of the transmission}

We give hereafter the main mathematical elements allowing to get the analytical formulae of the zero-order transmission though a grating-mode-resonant filter based on a subwavelength metallic slit grating, in TM-polarization.

Whereas the well-known Rigorous Coupled Wave Analysis is based on Fourier expansions of the electromagnetic fields and of the spatial distribution of the electrical permittivities, in every region of the system, we may resort to a true modal expansion of the field along the one-dimensional grating region, with some practical simplifications regarding boundary conditions \cite{Wirgin1984,Barbara2002}. By this way, we get a quite precise analytical formula giving the transmission in function of all opto-geometrical parameters, expandable to the perfect metal case. First, we only consider good (real) metals in the spectral range of interest, so that relevant surface impedance  conditions can be applied along metallic interfaces. Tangential fields obey the relation $\textbf{E}^{tan}\propto Z(\textbf{n}\times \textbf{H}^{tan})$ where $\textbf{n}$ is a vector normal to the metal surface, and $Z=1/\sqrt{\varepsilon}$, $\varepsilon$ being the metal relative permittivity ($|Z|\ll 1$). Secondly, we assume the vertical walls of the slit apertures as perfectly reflecting so that the field eigen modes are perfectly known (surface impedance could also be applied within the slits \cite{Locbilher1993,Leperchec2008} but the analytical calculation becomes a bit less explicit).

The incident light is transverse-magnetic (TM) polarized (magnetic field parallel to the slits, i.e. along the $y$-direction) and comes from the region $z<0$. $n_{i}, n_g, n_{t}$ are optical indices of the incidence, waveguide, and transmission regions, respectively. For sake of generality, we also consider the dielectric medium filling the grating slits as possibly different, with an index $n_d$. Then, the magnetic field transmitted outside the waveguide is written as a Rayleigh expansion:
\begin{equation}
H_y^{t}(x,z\geq h_g+h_{met})=\sum_ {m=-\infty}^{+\infty} T_m e^{jk n_{t}(\gamma^{t}_m x+\beta^{t}_m(z-h_g-h_{met}))}
\end{equation}
with $k=2\pi/\lambda$, $\beta_m^{t}=\sqrt{1-(\gamma_m^{t})^2}$, 
where each $\gamma_m^{t}$ respects the conservation of the tangential wave vectors $n_{i}\gamma_m^{i}=n_{g}\gamma_m^{g}=n_{t}\gamma_m^{t}$, with $n_{i}\gamma_m^{i}=n_{i}\sin\theta + m\lambda/P$, $\theta$ being the incidence angle. 

The total transmission in TM-polarization will be given by:
\begin{equation} 
Tr=\Re \left( \sum_m \frac{n_i\beta_m^{t}}{n_t\beta_m^{i}}|T_m|^2 \right) 
\end{equation}
As we use a sub-wavelength grating period ($n_{t}P<\lambda$) all over our spectral range of study, transmitted waves of high order are evanescent ($\beta_{m \ne 0}^{t}$ imaginary) and will not
contribute to far-field transmission. 

Applying then the matching conditions of the fields at every interface and using the 
\textit{one-mode} approximation ($w<<\lambda/n_d$) for the field inside the slit $h$-thick, we get (after long algebra):

\begin{equation}
T_m=\frac{(w/P) S_m }{(\beta_m^g/n_g)q_m+Z p_m}\Gamma(h)
\end{equation}
where
\begin{eqnarray}
S_m=\sec(kn_i\gamma^{i}_m w/2)\\
X_m=k n_g h_g\beta_m^g\\
Y_m=(n_g\beta_m^{t})/(n_{t}\beta_m^g)\\
 p_m =(1-j Y_n\tan(X_m))\cos X_m\\
 q_m=(Y_n -j\tan(X_m))\cos X_m
\end{eqnarray}

And
\begin{equation}\label{eq:Gamma}
 \Gamma(h)=\frac{[(n_d^{-1}-Z)(1-D_g^+)+(n_d^{-1}+Z)(1+D_g^-)] V} {(1+D_{i}^-)(1+D_g^-) \phi^{-1} -(1-D_{i}^+)(1-D_g^+)\phi }
\end{equation}
where $\phi=e^{jk n_d h}$, and
\begin{eqnarray}
 D_g^{\pm}=(\frac{1}{n_d}\pm Z)\frac{w}{P}\sum_ {m=-\infty}^{+\infty} \frac{S_m^2}{(q_m/p_m)(\beta_m^g/n_g)+Z}\label{eq:Dgreal} \\
 D_{i}^{\pm}=(\frac{1}{n_d}\pm Z)\frac{w}{P}\sum_ {m=-\infty}^{+\infty} \frac{S_m^2}{(\beta_m^{i}/n_i)+Z} \\
 V = \frac{2 S_0\cos \theta}{\cos\theta+n_dZ}
\end{eqnarray}

$V$ is an excitation term related to the incident plane wave. Subsequently, while $n_{t}P<\lambda$:
\begin{equation}
 Tr=\frac{n_{i}\beta_0^{t}}{n_{t}\cos\theta} \left|\frac{ (w/P)S_0 }{ q_0 (\beta_0^g/n_g)+ Z p_0}\Gamma(h) \right|^2
 \label{eq:Trsimplified}
\end{equation}

This analytical formula can rightly describe the behavior of the system even when Fabry-Perot-type resonances may occur within metallic apertures (i.e. when $n_c h \sim \lambda/2$). When $h_g=0$ and $n_i=n_t=1$, we get the transmission through a sub-wavelength slit grating made in a metallic screen and surrounded by air. In this work, we only focus us on cases of very thin gratings. 

For real metals ($Z\neq 0$ and $|Z| \ll 1$), (\ref{eq:Trsimplified}) is quantitatively reliable while $h$ remains greater than the metal skin depth $\delta$ (with $h>2\delta$ preferably), as 
the symmetric or anti-symmetric coupling of surface waves \textit{through} metallic parts is not described by this model. Also, the cavity width $w$ must not be extremely small because surface-plasmons coupling between vertical walls would become stronger and would modify the effective index of the waveguide mode inside the slit, with more optical absorption than predicted (mainly for deep slits and/or at visible frequencies).

\subsection{Perfect metal case}

Now, let $Z=0$. Eq.(\ref{eq:Gamma}) becomes:
\begin{equation} \Gamma (h)= \frac{V/n_d}{\cos(kn_ih) (D_{i}+D_g) - j\sin(kn_ih)(1+D_{i}D_g)}
\end{equation}

If we assume a normal incidence ($\theta=0$), then $S_0=1$, $V=2$, $\beta_0^g=\beta_0^t$=1, and:
\begin{equation}
 Tr=\frac{n_{i}}{n_{t}} \left |\frac{ 2 w n_g/(n_dP q_0)}{ [\cos(kn_ih) (D_{i}+D_g) - j\sin(kn_ih)(1+D_{i}D_g) ] }\right |^2
\end{equation}
with
\begin{eqnarray}
 D_g=\frac{1}{n_d}\frac{w}{P}\sum_ {m=-\infty}^{+\infty}\frac{n_g }{\beta_m^g}  \frac{ p_m }{q_m }\sec^2(m\frac{\pi w}{P})\\
 D_{i}=\frac{1}{n_d}\frac{w}{P}\sum_ {m=-\infty}^{+\infty} \frac{n_i}{\beta_m^{i}}\sec^2(m\frac{\pi w}{P})
\end{eqnarray}
When $kn_dh<<1$, $\cos(kn_dh)\approx 1$, $\sin(kn_dh)<<1$, and we get:

\begin{equation}
 Tr \approx\frac{n_{i}}{n_{t}} \left|\frac{ 2 }{(q_0/n_g) \sum_ {m=-\infty}^{+\infty} A_m } \right|^2
\end{equation}
where  $A_m$ are given by (\ref{eq:Am}).
Moreover, $A_m=A_{-m}$,  $Y_0=n_g/n_t$ and we have the following sequel
\begin{eqnarray}
A_0=n_i+p_0(n_g/q_0),  \\ 
q_0/n_g=\cos(kn_gh_g)/n_{t}-j\sin(kn_gh_g)/n_g, \\
p_0=\cos(kn_gh_g)-j\sin(kn_gh_g)(n_g/n_{t}).
\end{eqnarray}
Finally, for the perfect metal case, at normal incidence, and $kh<<1$, we find the zero-order transmission given by Eq.(\ref{eq:transm}).
By taking $w=P$ (no grating), all $A_m=0$, we retrieve the transmission response of a simple homogeneous slab of index $n_g$, with maximum $Tr=1$ when $\sin(kn_gh_g)=0$ and minimum transmission when $\cos(kn_gh_g)=0$.

\subsection{Resonance conditions}

The terms $A_m$ (\ref{eq:Am}) are complex quantities whose moduli decrease as $m\rightarrow \infty$. Resonances may occur when the imaginary part of the denominator in (\ref{eq:transm}) cancels, say:

\begin{equation}
\left(\frac{n_g}{n_t}+\frac{n_{i}}{n_g}\right)\tan (kn_g h_g) \approx \frac{2}{n_t} \Im \left(\sum_{m >0}A_m\right)
\end{equation}

In particular, once the first waveguide mode becomes propagating inside the slab, i.e. $\beta^g_1>0$, $A_1$ becomes entirely imaginary, so we can directly get an approximated resonance condition for the strong TM-transmission peak we are studying in this work, by neglecting high orders terms $m\geq 2$. However, neglecting high orders is only relevant for rather large values of $w$, i.e. when the factors $\sec^2(m\pi w/P)_{m>1}$ in (\ref{eq:Am}) have weak amplitudes. As $w$ decreases, we have to take into account more and more $A_m$ in the resonance condition hereabove.
Note that other resonances can also appear at oblique incidence, related to anti symmetrical surface waveguide modes, but this will not be discussed in the present paper.

\subsection{Transmission nulls}

Let us come back to the real metal case (\ref{eq:Trsimplified}). We can see that transmission may cancel when (\ref{eq:Dgreal}) diverges ($\Gamma(h) \rightarrow 0$), i.e. when the following condition is fulfilled:
\begin{equation}
q_m(\beta_m^g/n_g)+Z p_m\approx 0
\end{equation}
which reduces to $j\tan(X_m)=Y_m$ in the perfect metal case. Zeros of transmission might also occur when $\beta_m^i/n_i+Z=0$ (divergence of $D_{i}$), corresponding to classical surface-plasmon excitations between the grating and the incidence region, but such resonances do not happen here as $P<<\lambda$.


\end{document}